\documentclass[fleqn,12pt]{article}

\usepackage{amsmath,amsfonts,amssymb,eucal,graphicx,graphics,bm}
\usepackage[sort&compress,numbers]{natbib}
\usepackage{epsfig}
\usepackage{wrapfig}
\usepackage{subfigure}
\usepackage{amssymb}
\usepackage{epsfig}
\usepackage{subfigure}
\usepackage{graphicx}
\usepackage{color}
\usepackage[dvipsnames]{xcolor}
\usepackage{textcomp}
\usepackage{url}
\usepackage{float}
\usepackage{bm}
\usepackage{authblk}
\usepackage[utf8]{inputenc}
\usepackage[T1]{fontenc}
\usepackage[toc,page]{appendix}
\usepackage[parfill]{parskip}

\title{Scaling of Urban Income Inequality in the United States}

\author[1, 2]{Elisa Heinrich Mora}
\author[3, 2]{Jacob J. Jackson}
\author[4, 2]{Cate Heine}
\author[2]{Geoffrey B. West}
\author[2]{Vicky Chuqiao Yang}
\author[2]{Christopher P. Kempes}

\affil[1]{Minerva Schools at KGI, San Francisco, CA 94103, USA}
\affil[2]{Santa Fe institute, Santa Fe, NM 87501, USA }
\affil[3]{Brown University, Providence, RI 02912, USA}
\affil[4]{Massachusetts Institute of Technology, Cambridge, MA 02139, USA}

\begin{document}
\maketitle

\begin{abstract}
Urban scaling analysis, the study of how aggregated urban features vary with the population of an urban area, provides a promising framework for discovering commonalities across cities and uncovering dynamics shared by cities across time and space. Here, we use the urban scaling framework to study an important, but under-explored feature in this community---income inequality. We propose a new method to study the scaling of income distributions by analyzing total income scaling in population percentiles. We show that income in the least wealthy decile (10\%) scales close to linearly with city population, while income in the most wealthy decile scale with a significantly superlinear exponent. In contrast to the superlinear scaling of total income with city population, this decile scaling illustrates that the benefits of larger cities are increasingly unequally distributed. For the poorest income deciles, cities have no positive effect over the null expectation of a linear increase. We repeat our analysis after adjusting income by housing cost, and find similar results. We then further analyze the shapes of income distributions. First, we find that mean, variance, skewness, and kurtosis of income distributions all increase with city size. Second, the Kullback-Leibler divergence between a city's income distribution and that of the largest city decreases with city population, suggesting the overall shape of income distribution shifts with city population. As most urban scaling theories consider densifying interactions within cities as the fundamental process leading to the superlinear increase of many features, our results suggest this effect is only seen in the upper deciles of the cities. Our finding encourages future work to consider heterogeneous models of interactions to form a more coherent understanding of urban scaling.
\end{abstract}


\section{Introduction}
Throughout human history, the global urban population has grown continuously. More than half of the global population is currently urbanized, placing cities at the center of human development \cite{united2018}. It is estimated that by 2030, the number of megacities, cities with more than 10 million inhabitants, will increase from 10 to approximately 40 \cite{united2018}. Thus, there is an urgent need for a quantitative and predictive theory for how larger urban areas affect a wide variety of city features, dynamics, and outcomes \cite{lobo2020urban}. Perhaps most critically, we need this theory to address how larger cities positively and negatively affect socioeconomic outcomes and the quality of life of individuals. 

Previous research has demonstrated power-law-like relationships between urban population (also referred to as size later in the text)  and many urban features such as GDP, patents, crime, and contagious diseases that persist globally \cite{bettencourt2007growth, bettencourt2016urban, meirelles2018evolution, zund2019growth, sahasranaman2019urban}. These relationships can often be described by 
\begin{equation}\label{eq:scaling}
    Y = Y_0 N^{\beta}\;,
\end{equation}
where $Y$ is an urban feature, such as GDP or number of crime instances, $N$ is the population of the city, $Y_0$ is a constant, and $\beta$ is the scaling exponent. For many urban outputs, the scaling exponent $\beta$ is greater than 1, suggesting greater rates of productivity (in both the positive and negative sense) in more populated cities. These observations, known as urban scaling, suggest that a small set of mechanisms significantly influence a variety of urban features across diverse cities \cite{bettencourt2013origins, yang2019modeling}. Understanding these mechanisms has important implications for developing more prosperous and safer cities. In this framework, desirable aspects with $\beta>1$ have positive returns to scale, while desirable aspects with $\beta<1$ have a less than linear return to scale, demonstrating a diseconomy of scale. Similarly, for undesirable features $\beta>1$ shows a diseconomy of scale since the associated per-capita costs would be increasing with city size.

One important aspect of urban features that remains under-explored in the urban scaling framework is economic inequality. Inequality has fundamental implications for individuals' quality of life and the productivity and stability of societies \cite{bowles2012new}. Past research has heightened debate about economic inequality and its relationship with economic growth and general welfare \cite{piketty2014capital, tabelline, alesina, lizou, forbes, barro, pagano}. Many have raised concern of its negative effects on political stability \cite{alesina1996income, piketty2014capital}, crime \cite{kelly2000inequality} and corruption \cite{jong2005comparative}. It has been shown that more unequal places have higher murder rates, grow more slowly, and the correlation between area-level inequality and population growth is positive \cite{NBERw14419}. Economic inequality is usually measured in terms of the dispersion in the distribution of income or wealth, such as in the Gini Coefficient. Some past research has noted larger cities are correlated with increasing Gini coefficient in income distribution \cite{survivalofthefittest,cottineau2016defining,baum2013inequalityandcitysize} , but it remains unclear if there are systematic relationships between other features of the income distributions and urban area size. Furthermore, characterizing distributions by a single metric may lose important information \cite{piketty2014capital} -- for example, does being poor in bigger cities correspond to a higher or lower standard of living than being poor in a smaller city?

\begin{figure}[!htb]
    \centering
    \includegraphics[width=0.9\textwidth]{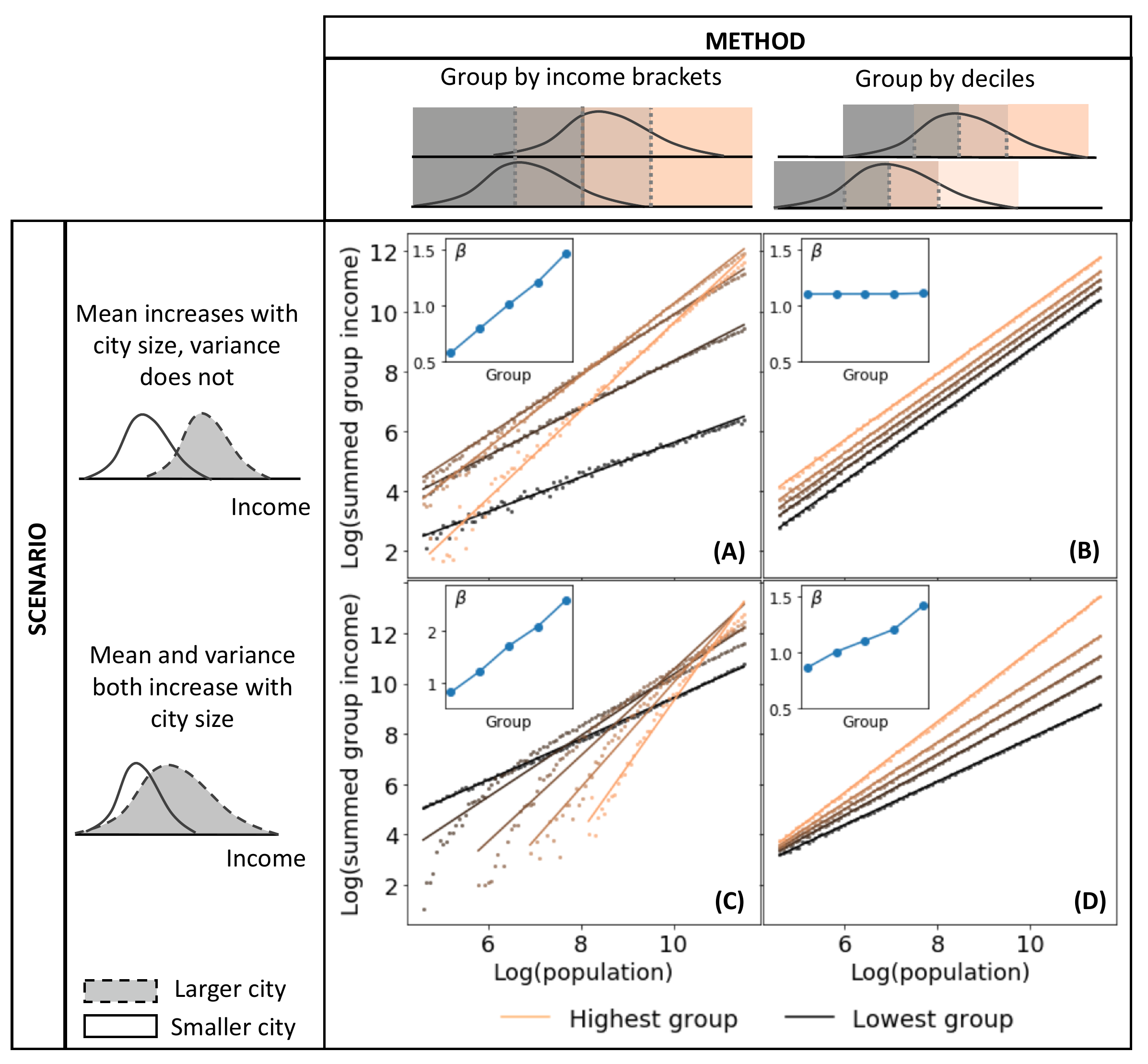}
    \caption{Illustration comparing two methodologies---scaling obtained from grouping by income bracket (A and C) and that by decile (B and D). Using simulated log-normal income distributions in two scenarios---log-mean increases with city size while log-variance remains the same (A and B), and log-mean and log-variance both increases with city size (C and D). The income distributions are illustrated on a log-scale. The income-bracket grouping (A and C) leads to differences in the groups' income scaling for both scenarios, and fails to distinguish whether larger cities have more dispersion in their income distributions. The decile grouping (B and D) leads to the differences in the groups' scaling observed only when the dispersion increases with the population. The insets show how scaling exponent ($\beta$) varies with income groups (bracket or decile).
}
    \label{fig:method_cartoon}
\end{figure}

A few recent studies \cite{sarkar2018scaling, sarkar2019urban} have investigated the scaling of total income in various income brackets in Australia. These studies find that the total income in lower income brackets scales sublinearly or linearly, while higher brackets scale superlinearly, suggesting greater income agglomeration in the higher income categories in more populated cities. While these studies are informative and provide a new measure for inequality in terms of absolute income (instead of relative income, as in the Gini Coefficient), a limitation is that this measure confounds inequality with average income, which increases with city population. In particular, the ``equal'' situation in this new measure of inequality is when the total income for all income brackets scales linearly. However, given that total income scales superlinearly in cities globally \cite{bettencourt2007growth, bettencourt2016urban}, this ``equal'' situation is unlikely to occur. For example, even if the shapes of income distributions remain identical, income bracket aggregations follow distinct scaling relationships as a result of differences in mean. Figure~\ref{fig:method_cartoon} A and C illustrates this behavior using simulated log-normal distributions. While the measure of inequality proposed in \cite{sarkar2018scaling, sarkar2019urban} can be valuable for some applications, it would be useful to untangle the increase in mean from the greater dispersion in income. 

In this manuscript, we address a few keys questions:  (1) How does income inequality (adjusted for shifting average income) systematically change with city size? (2) How different is the income of rich and poor people (measured by percentiles of the population) in small and large cities, and how does this difference scale with city size? (3) Are poor people in a larger city better off than poor people in a small city, after adjusting by the cost of living? How about the same for rich people?

Here, we propose a new method to study the scaling of inequality by analyzing total income scaling in population percentiles. We show that income in the least wealthy decile (10\%) scales almost linearly with city size, while that in the most wealthy decile scales with a significantly superlinear exponent. This illustrates that the benefits of larger cities are increasingly unequally distributed, and for the poorest income deciles, city growth has no positive effect on income growth over the null expectation of a linear increase. 
We then introduce systematic considerations of the entire distribution of income to show which income distribution features are changing with city size. We find that the mean, variance, skewness, and kurtosis of the income distribution all scale systematically with city size. We introduce a KL-divergence procedure to systematically compare all moments and find that comparisons with the largest cities also demonstrate a systematic scaling with city size, indicating that the overall shape of income distribution is radically shifting with city size. We then attempt to identify actual changes in purchasing power with city size by normalizing income by housing costs, which also grow superlinearly with city population. Finally, we discuss how these observations can be connected with the proposed mechanisms underlying urban scaling.

\section{Data and methods} \label{sec:empirical_incomes}

\subsection{Scaling of income aggregated by deciles}

We propose a new method to investigate the scaling of income aggregated by deciles in each city (i.e., the bottom 10\%, the next 10\%, and so on). The number of individuals in decile $n$ of city $i$ is, $N^{(n)}_{i} = N_i/10$, where $N_i$ is the population of city $i$. The total income in decile $n$ of city $i$, $Y^{(n)}_{i}$ is, 
\begin{equation}
Y^{(n)}_{i} = \sum_{j \in \textbf{D}^{(n)}}y_{i, j}\;,
\end{equation}
where $\textbf{D}^{(n)}$ are the individuals in income decile $n$, and $y_{i,j}$ is the income of individual $j$ in city $i$. See Supplemental Materials for more details on the decile assignment in our computational implementation.

Figure~\ref{fig:method_cartoon} C and D illustrate this method on simulated log-normal income distributions. Panel C represents the situation in which cities shift in log-mean with city size, but do not shift in log-standard deviation, and panel D represents the situation in which cities increase both log-mean and log-standard deviations with city size. We consider the former case an example of the ``equal'' situation, and this method should lead to no variation in scaling exponents across deciles. Variations in scaling exponent only occur for the latter case. We also contrast the results of our method with that of the grouping by income bracket method in Figure~\ref{fig:method_cartoon} A and C, where variations in scaling exponents occur for both scenarios.

\begin{figure}[!h]
\centering
\includegraphics[width = 0.5\textwidth]{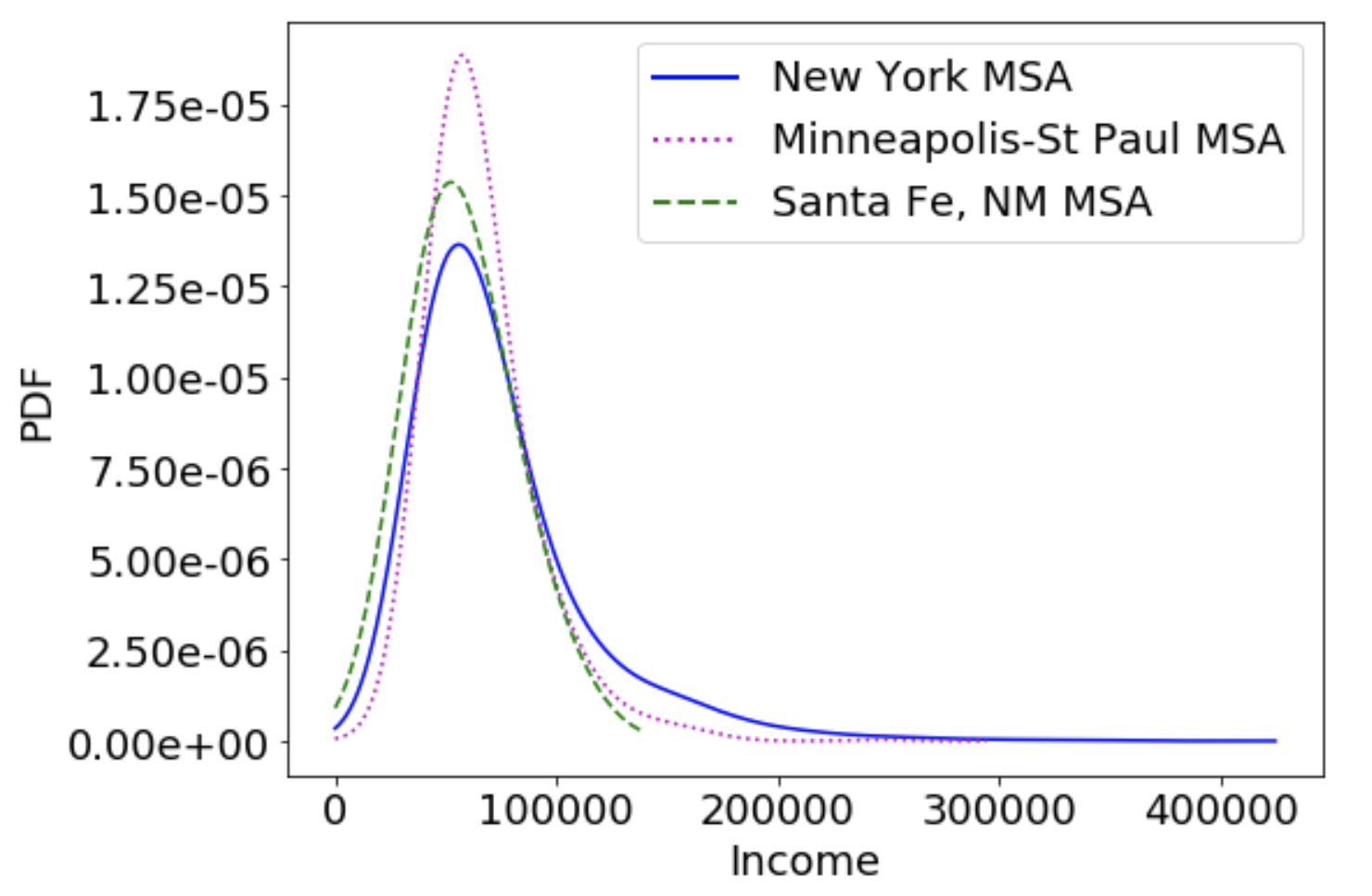}
  \caption{Examples of the estimated income distributions using census tract data. Income is measured in US dollars. The three metropolitan areas shown are: New York-Newark-Jersey City, NY-NJ-PA, population 20,316,622; Minneapolis-St. Paul-Bloomington, MN-WI, population 3,670,397; Santa Fe, NM, population 204,396.}
  \label{fig:KDE_pdfs}
\end{figure}

\subsection{Data and income distribution estimation}
The primary dataset used in our analysis is the 2015 American Community Survey conducted by the US Census Bureau (see Supplementary Materials for more detail). We use the income data reported on the level of census tracts, small local areas of on average 4500 people, of which on average 2300 reported income. We infer the individual-level income distribution in Metropolitan Statistical Areas (MSAs) by applying the Gaussian kernel density estimator with a widened Silverman bandwidth function on the census-tract-level data. This method assumes income in each census tract is distributed as a Gaussian. The mean equals the average income of the census tract, and the standard deviation is calculated as a function of the number of data points. Aggregating the Gaussian probability density functions (PDFs) for each census tract in the MSA produces an estimated income PDF for the MSA. Examples of the estimated individual-level income distribution for a few MSAs are shown in Figure~\ref{fig:KDE_pdfs}. 

\subsection{Analysis of income scaling in deciles}
The estimated income distributions for US cities are grouped into deciles: the 10\% of the population which reports the lowest income is grouped into the first decile (decile \#1), and likewise for all ten deciles up to the 10\% of the population which has the highest income (decile \#10). We then estimate the scaling exponent of total income for all deciles. We estimate the scaling exponent, $\beta$, and corresponding confidence intervals, by performing an ordinary least square regression of the log-transformed variables, $\log(Y^{(n)}_i) = \beta \log(N^{(n)}_i) + c$, and $\beta$ and $c$ are the fitted parameters. This methodology is consistent with previous research such as \cite{bettencourt2007growth}.

\subsection{Analysis of distributions}
We further analyze how the shapes of the income distributions vary with city population. We first compute the first four statistical moments, mean, variance, skewness, and kurtosis, for income distributions of each city, and analyze how they vary with population. We then compute the Kullback-Leibler (KL) divergence between each city's income distribution and that of the largest city (New York-Newark-Jersey City MSA). The KL divergence measures how different one distribution is from another, while the zero value indicates the two distributions are identical, and a greater value indicates more divergence. Mathematically, the KL divergence between two discrete distributions of random variable $x$, $P(x)$ and $Q(x)$ is, 
\begin{equation}
    KL(P||Q) = \sum_x P(x) \log(P(x)/Q(x))\;.
\end{equation}
\subsection{Adjusting income by housing cost}
In order to normalize income by the cost of living, we calculate total housing cost in a census tract as $cost = 12 \;(u_{rent}  r + u_{own} \; o)$, where the average monthly rent $r$, the average monthly owner costs $o$, and the number of units of each type $u_{rent}$ and $u_{own}$ are all taken from the 2015 American Community Survey (see Supplementary Materials for more detail and access information). We then repeat the decile-grouped analysis on income adjusted for housing cost, as well as analyze how the proportion of income spent on housing varies with city size in each decile. 

\section{Results}
\subsection{Scaling of income in deciles}
\begin{figure}[!htb]
  \centering
  \includegraphics[width=0.8\textwidth]{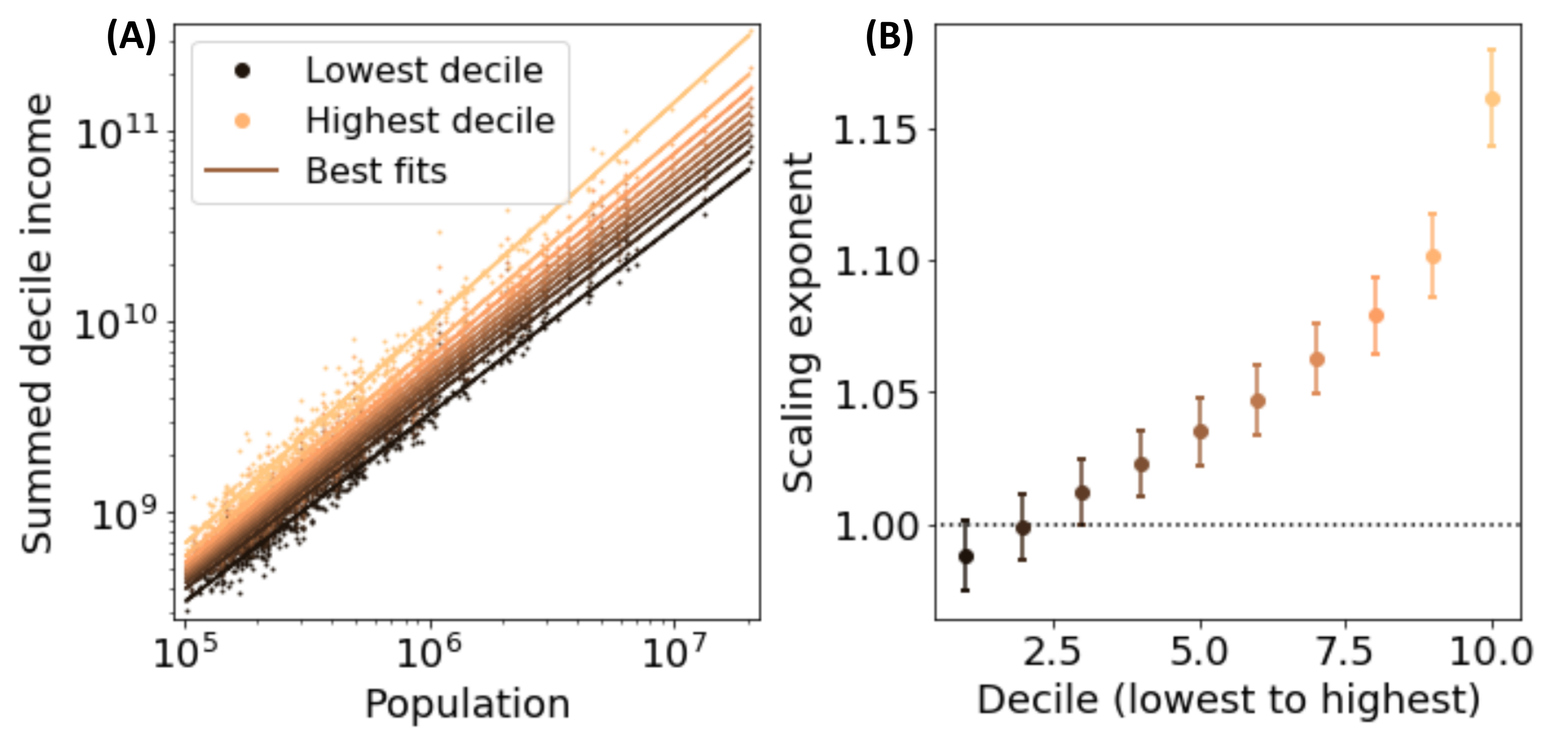}
    \caption{Scaling of income (in US Dollars) by population for deciles of US MSAs. (A) Scaling of total income in deciles (B) Scaling exponents ($\beta$) of each decile and corresponding 95\% confidence intervals. The dashed line is $\beta = 1$ to help guide the eye. Higher-income deciles exhibit greater scaling exponents than lower income deciles, and the lowest deciles exhibit near-linear scaling. The scaling exponents for aggregated income in city, combining all deciles, is 1.07.}
    \label{fig:decilescaling}
\end{figure}

The results for scaling of income aggregated in deciles are summarized in Figure~\ref{fig:decilescaling}. For the lowest two deciles, the scaling exponent $\beta$ is linear or slightly sublinear ($0.97$). For upper deciles, $\beta$ is consistently superlinear, as high as $1.16$ as compared to the scaling exponent of total income in our dataset, $\beta = 1.07$. This shows that scaling effects are not equivalent for all segments of the population. The poorest two deciles in bigger cities make about the same income as their counterparts in smaller cities, while the wealthiest eight deciles in bigger cities make more than their counterparts in smaller cities, where the difference increases with the decile.

\subsection{Analysis of income distribution characteristics}
We further analyze how income distributions vary with urban area population by studying the statistical moments of the income distributions. We first examine the first four moments: mean, variance, skewness, and kurtosis.



\begin{figure}[htb]
\centering
\includegraphics[width = 1\textwidth]{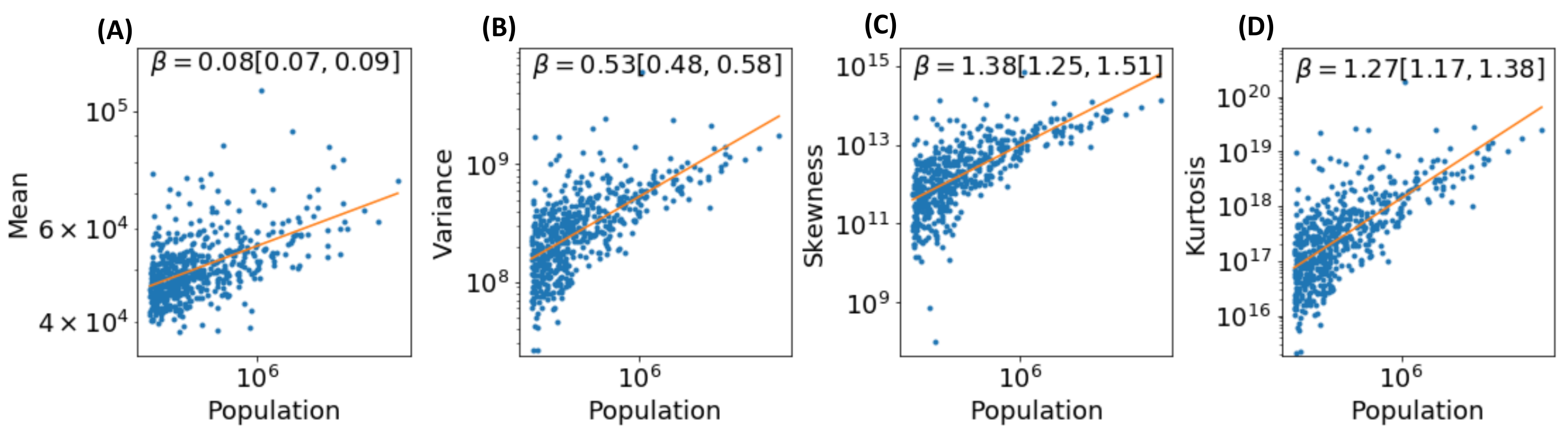}
  \caption{First four statistical moments of the estimated income distributions as a function of city size. The texts in each panel display the scaling exponent, $\beta$, and in the bracket, corresponding 95\% confidence intervals.}
  \label{fig:moment_dists}
\end{figure}

The scaling of the four moments of the estimated individual income distribution for all cities in our data is shown in Figure~\ref{fig:moment_dists}. The first moment, the mean, shows the well-characterized urban agglomeration effect: per-capita income increases with city size \cite{bettencourt2007growth}. The second, and third moments both increase similarly with city population, suggesting a widening of the distribution and increasing asymmetry with greater urban population. This can also be qualitatively observed in the example distributions in Figure \ref{fig:KDE_pdfs}. Lastly, the kurtosis also increases with population size, showing an increasingly heavy tail with greater urban population. 

We find a stronger relationship for higher statistical moments, indicating that for larger American cities, there is a more evident increase in the third and fourth moments. This means that there is a stronger increase in the growing tail of the distribution, in comparison to the first two statistical moments. This gives us an interesting indication of the distribution of economic benefits.

\begin{figure}[htb]
\centering
\includegraphics[width=0.5\textwidth]{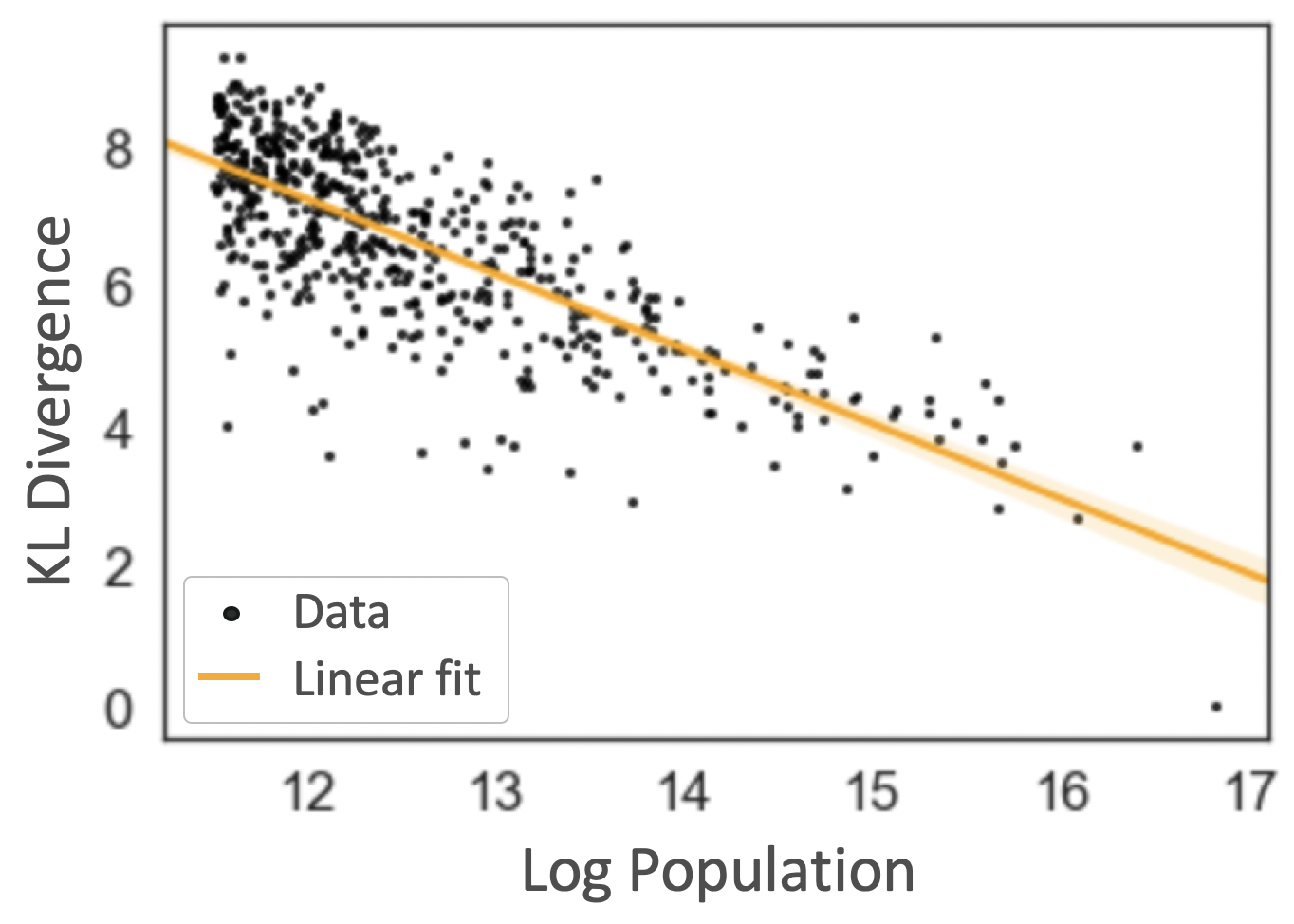}
\caption{Kullback-Leibler divergence between the estimated income distributions and that of the largest city, as a function of log population. The Spearman correlation is -0.718.}.
\label{fig:klvalues}
\end{figure}

Another useful perspective on the scaling of the income distributions is to compare large and small cities using measures that consider the entire distribution through the KL divergence. Figure \ref{fig:klvalues} shows the KL divergence between each US city and the largest city, as a function of the log-transformed city population. The KL divergence, in general, decreases with city population, and approaches zero as the population approaches that of the largest city. This behavior suggests that as cities get smaller, their income distributions are increasingly dissimilar to that of the largest city. The Pearson correlation between the two variables in Figure~\ref{fig:klvalues} is  $-0.259$, while the Spearman correlation is $-0.718$. The Pearson correlation measures the linear correlation between two variables, while the Spearman correlation measures the rank correlation, and assesses how well relationship between two variables can be described by a monotonic function, regardless of linearity. This finding suggests that population and the KL divergence tend to change together, but not necessarily at a constant rate. While we can identify a general scaling trend, our data also exhibit frequent outliers and deviations.


\subsection{Scaling of decile income adjusted by housing cost}
\begin{figure}[!h]
    \centering
    \includegraphics[width = 0.9\textwidth]{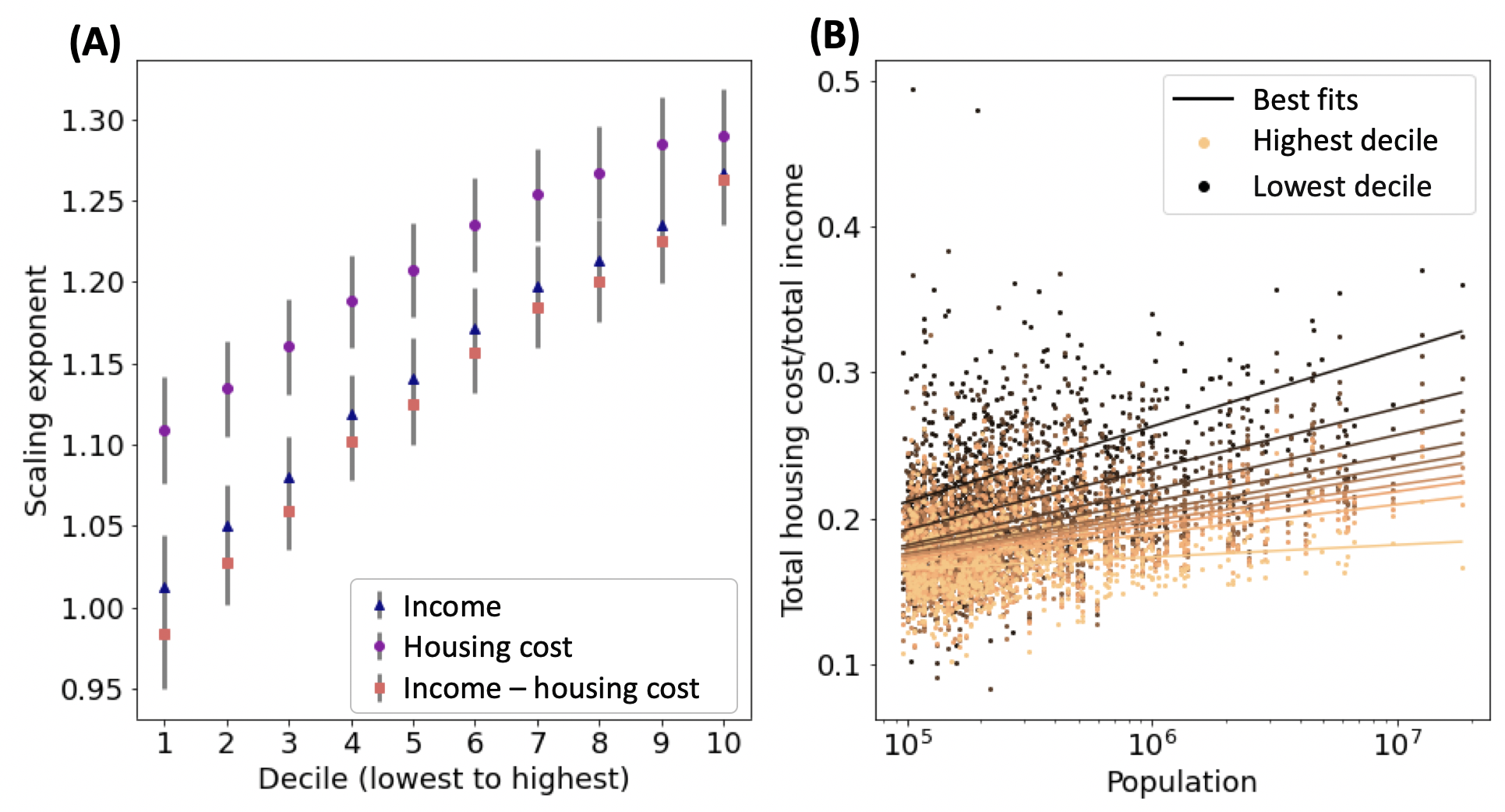}
    \caption{Comparing the scaling of housing cost and income. (A) The scaling of total income, total housing cost, and the difference between total income and housing cost, for each decile. Housing cost scales with greater exponents than income for all deciles. The housing-adjusted income exhibits similar variation across deciles as total income. (B) Ratio between housing cost and household income as a function of city population. In the poorest deciles (dark brown), the proportion of income spent on housing increases sharply with city size; in the wealthiest deciles (orange), this proportion remains stagnant.
    }
    \label{fig:housing_cost_pct}
\end{figure}

While the differences in income scaling that we have identified are important, they are not necessarily grounded in differences in the experiences of urban residents---cost of living can vary drastically across and within US cities, and if cost of living is changing in the exact same way as income, differences in income scaling between groups begin to lose meaning. In order to understand whether the differences in income scaling we see between deciles create differences in affordability and purchasing power, we look at changes in housing cost with city size. 
By analyzing, in combination, aggregate household income and aggregate housing cost for each census tract, we find that aggregate housing cost scales faster than aggregate income for \textit{every} decile, implying that while income per person increases with city size, larger cities may still be overall less affordable. This difference is more dramatic for the poorer deciles---in the bottom decile, housing cost scales with $\beta = 1.11$ while income scales with $\beta = 1.01$; in the top decile, housing cost scales with $\beta = 1.29$ while income scales with $\beta = 1.27$. This is visualized in Figure~\ref{fig:housing_cost_pct}A---income exponents begin to catch up to housing cost exponents in richer deciles, but never as high as housing cost. Perhaps more intuitively, in Figure \ref{fig:housing_cost_pct}B, we can see that the ratio between total housing cost and total income grows with city size for every decile, but more dramatically for poorer deciles. Together, these results imply a widening gap between richer and poorer residents in affordability of cities with city size.

\section{Discussion}

\begin{figure}[htb]
\centering
\includegraphics[scale=0.4]{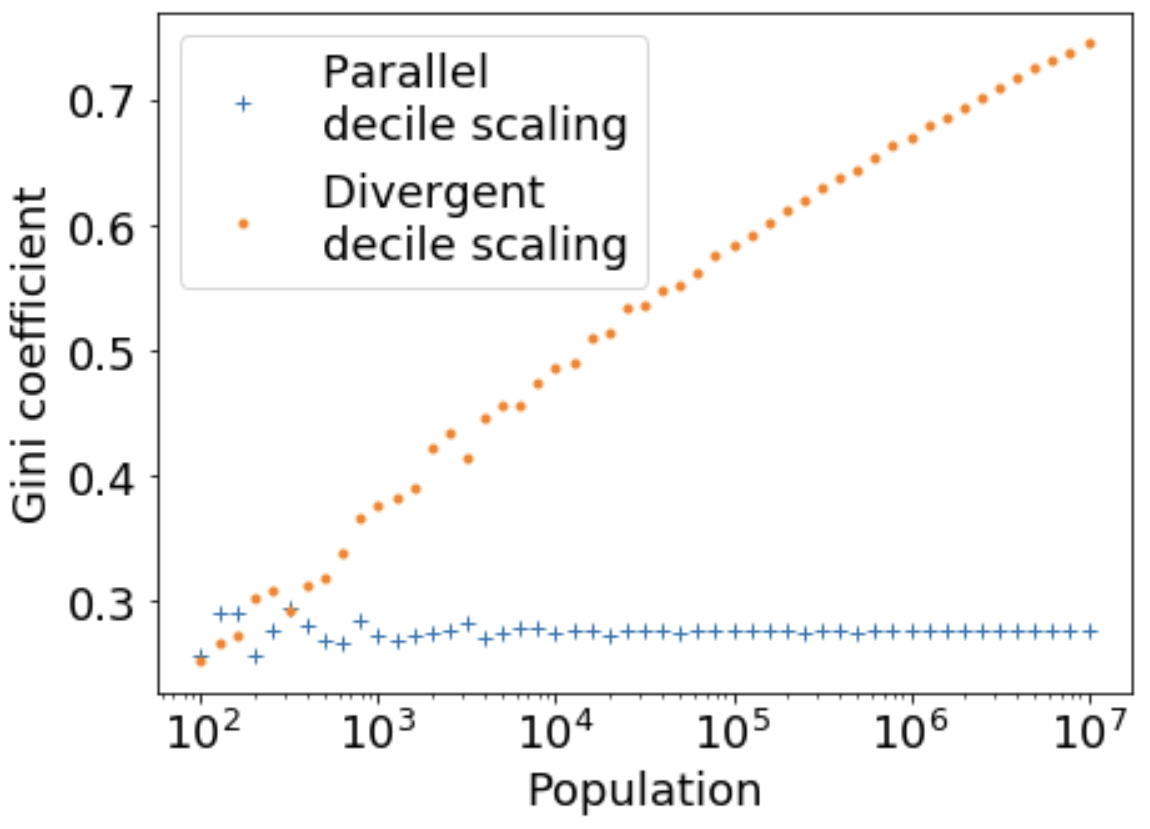}
\caption{Changes in the Gini coefficient with urban population in simulated log-normal distributions. For a scenario of parallel decile scaling (Figure~\ref{fig:method_cartoon}B) and for a scenario where the deciles have divergent scaling (Figure~\ref{fig:method_cartoon}D). As expected, the divergent scaling observation corresponds to increasing Gini coefficient with population.}
\label{fig:gini}
\end{figure}

Here we proposed a new method to study the scaling of income distributions and income inequality in urban areas. The aggregated income in income deciles scale systematically with city size. The bottom decile scales with an exponent slightly below $1$ and the top decile with an exponent of $\beta=1.15$. This result suggests that the benefits of larger cities are increasingly unequally distributed, and for the poorest income deciles, cities have no positive effect over the null expectation of a linear increase. Much has been written about the apparent increasing gains of large cities \cite{bettencourt2007growth, bettencourt2016urban}, such as greater GDP, higher wages, and more patents per capita. Our results show that the increasing benefits of city size are not evenly distributed to people within those cities. 
We further show systematic variations in distribution characteristics. Besides greater mean, distributions of bigger cities also exhibit greater spread, greater asymmetry, and heavier tail. These perspectives can be explicitly connected to traditional measures of income inequality, such as the Gini coefficient. Like the Gini Coefficient, our method characterizes the overall dispersion of income distributions (see Figure~\ref{fig:gini}), but it also provides more detailed information that is not characterized by Gini, such as how the urban agglomeration effect alters the incomes of relatively poor or rich people differently.



Although our results appear to closely align with those of Sarkar et al.\cite{sarkar2018scaling, sarkar2019urban}, which analyze Australian income data, the difference in methodology (aggregation by income brackets vs.~by deciles) should lead to different interpretations of the scaling exponents derived. In particular, the baseline ``equal'' situation is different in the two methods---in Sarkar et al., when total income in all income brackets scale with the linear exponent, and in our methods, when total income in all deciles scale with the same exponent (either linear or nonlinear). 

Our paper offers new contributions to the literature. First, we develop a new method to study income inequality in the urban scaling framework, which untangles the systematic shift in mean from the study of income inequality. This method enables us to study how income agglomeration effects vary between relatively rich and poor people, after accounting for the systematically increasing mean with population size. Second, our analysis including housing cost demonstrates that despite agglomeration effects on income, bigger cities are less affordable for people of all deciles in the sense that they spend proportionally more of their income on housing; this is especially true for lower-income people. Third, our analysis extends beyond the single-parameter characterization of income inequality. We analyze more complex properties of income distributions through analyzing statistical moments and KL divergence, and reveal systematic variations with city size. Fourth, our results suggest new directions for understanding mechanisms of urban agglomeration effects---it is important to extend beyond theories considering homogeneous densifying interactions to those which account for heterogeneity. 


Understanding the underlying mechanisms of why inequality is systematically scaling with city size is of great future interest with many potential implications. Urban scaling theory in general proposes densifying interactions within cities as the fundamental process leading to the superlinear increase of many features \cite{bettencourt2013origins, yang2019modeling, gomez2016explaining, pan2013urban}. Our analysis shows that the superlinear scaling is not seen within all subsections of the city. The superlinear scaling of total wealth is driven by the top income deciles, and is not matched proportionally by the lowest deciles. This adds another dimension to considerations of the underlying mechanisms of urban scaling theory: what processes are leading to the increasingly unequal distribution of wealth in larger cities? We explored the idea of city heterogeneity as an indirect proxy for heterogeneous interaction rates. One hypothesis of the mechanism driving superlinear scaling of income with city size is that larger cities foster more and more diverse social and economic interactions, creating opportunities for the exchange of ideas and resources. Existing literature credits superlinear growth of income in cities to more opportunities for social contacts and interactions in large cities \cite{bettencourt2007growth, bettencourt2013origins}. Increased social contact with city size has been empirically confirmed \cite{bettencourt2014interactions}, and ties between individual's exposure to diverse social connections and economic outcomes have been shown empirically as well \cite{chetty}. Together, this seems to suggest that cities that are better mixed, allowing diverse parts of the population to be exposed to one another, should be overperforming with respect to urban scaling. We hypothesize that cities with high levels of economic segregation, inhibiting mixing between diverse populations, will underperform with respect to income scaling. Our finding encourages future work to consider heterogeneous models of interactions, as those clustered in space or social/work circles, to form a more coherent understanding of urban scaling.

	\setcounter{figure}{0}
	\setcounter{section}{0}
	\renewcommand\thefigure{S\arabic{figure}}  
	\renewcommand\thetable{S\arabic{table}}  
    
	\newpage
	\setcounter{equation}{0} 
	\setcounter{page}{1}
	
	\begin{center}
		\Large{Supplementary Materials for \\Scaling of Urban Income Inequality in the United States
		}
	\end{center}
	
	\normalsize
	
	\setcounter{equation}{0}
	\renewcommand\theequation{S\arabic{equation}}

\section{Data sources and methods}\label{sec:SM:data}

Income and housing cost data are from the 2015 American Community Survey, openly available through the United States Census Bureau at \url{https://www.census.gov/programs-surveys/acs}. The definition of cities we employ is the Metropolitan Statistical Areas (MSAs) defined by the United States Office of Management and Budget.

In our income measure in the decile scaling and distribution analysis (Sections 3.1 and 3.2), we multiply mean earnings by the total number of workers aged 16 years and older from census table S2001: Earnings in the Past 12 Months. As our income measure in the housing cost analysis (Section 3.3), we multiply the mean household income from census table S1901 by the total number of households for which we have rent and owner cost information. We calculate total housing cost in each census tract as $cost = 12 \;(u_{rent}  r + u_{own} \; o)$,
where average monthly rent $r$, average monthly owner costs $o$, and the number of units of each type $u_{rent}$ and $u_{own}$ are taken from census table DP04: Selected Housing Characteristics.

We provide an example of our division of census tracts into deciles in Table \ref{dec10140}, which depicts the census tracts in deciles 1 and 2 for Aberdeen, Washington. The process is as follows:
\begin{enumerate}
    \item For each MSA, sort census tracts by descending mean income.
    \item Iterate through census tracts in order of income, adding census tracts to decile 1 until adding the next census tract would push decile 1's population over 10\% of the total MSA population.
    \item Add the proportion of that census tract's population that would put decile 1 at exactly 10\% of the total MSA population and the same proportion of the census tract's income to decile 1; add the remaining population and income to decile 2.
    \item Repeat for all deciles, until there are 10 deciles with equal population (but increasing income).
\end{enumerate}

\begin{table}[htb]
\caption{An example illustrating our method of assigning deciles to census-tract data. The urban area in this example is Aberdeen, Washington. Census tracts 53027001200 and 53027001000 are each split between two deciles.}
{\footnotesize
\begin{tabular}{lrrrrrrrr}
\hline \\
 &  MSA &  MSA &  &&  CT mean  &  CT total  &  CT population &  Decile  \\
&  ID &   population & Decile  &      CT  ID &  income &  population & (this decile) &   population \\
\hline \\
0 &   10140 &          102471 &1&  53027001600 &           36011 &               6547.0 &                       6547.0 &            10247.1 \\
1 &   10140 &          102471 &1&  53027001200 &           37891 &               4514.0 &                       3700.1 &            10247.1 \\
2 &   10140 &          102471 &2&  53027001200 &           37891 &               4514.0 &                        813.9 &            10247.1 \\
3 &   10140 &          102471 &2&  53031950702 &           39852 &               4416.0 &                       2208.0 &            10247.1 \\
5 &   10140 &          102471 &2&  53049950300 &           40336 &               3613.0 &                       3613.0 &            10247.1 \\
6 &   10140 &          102471 &2&  53027940000 &           40611 &               1106.0 &                       1106.0 &            10247.1 \\
7 &   10140 &          102471 &2&  53045960100 &           41041 &               1396.0 &                        698.0 &            10247.1 \\
9 &   10140 &          102471 &2&  53027001000 &           41552 &               3544.0 &                       1808.2 &            10247.1 \\

\hline 

\end{tabular}
}
\label{dec10140}
\end{table}

\end{document}